\documentclass[twocolumn,superscriptaddress,amssymb,amsmath,a4paper]{revtex4}

\newif\ifAPLsubmission

\APLsubmissionfalse

\ifAPLsubmission
\else
\fi

\usepackage{graphicx}% Include figure files
\usepackage[usenames]{color}

\usepackage{dcolumn}% Align table columns on decimal point
\usepackage{bm}% bold math

\usepackage[germanb,english]{babel}
\usepackage[utf8x]{inputenc}

\newcommand{\etal}{\mbox{\it et~al.}}

\newcommand{\Brf}{\ensuremath{B_{\rm 1}}}
\newcommand{\Bfb}{\ensuremath{B_{\rm fb}}}
\newcommand{\B}{\ensuremath{B_{\rm 0}}}
\newcommand{\omegarf}{\ensuremath{\omega_{\rm rf}}}
\newcommand{\omegarfavg}{\ensuremath{\omega_{\rm com}}}
\newcommand{\Dfi}{\ensuremath{\Delta\varphi}}
\newcommand{\Hz}{\ensuremath{\mathrm{Hz}}}

\newcommand{\Tesla}{\ensuremath{\mathrm{T}}}

\newcommand{\fTHz}{\ensuremath{\fT/\sqrt{\Hz}}}

\newcommand{\fT}{\ensuremath{\mathrm{f}\Tesla}}

\def\Captionone{Perspective drawing of the 25-sensor 19-channel cMFM system
  installed in our Al shielding room in Fribourg.  The bottom plane
  holds the 19 measurement channels, while the mid plane and top plane
  hold three sensors each, used for field and gradient fluctuation
  suppression.}
\def\Captiontwo{Time series showing (averaged) MCG traces of a healthy
  subject for each of the 19 sensor positions.  Each trace is 1~s long
  and all graphs have the same (280~pT) vertical extension. For this measurement
  only a single (centered) reference sensor was used in the middle plane.}
\def\Captionthree{(Color online) Magnetic field map for the out-of-chest
  component of the cardiac field.  Crosses show sensor
  positions, and the superimposed trace (bottom center trace from
  Fig.~\protect\ref{fig:trace}) identifies, via the black dot, the
  time for which the map was evaluated.  Left (right): maps at the
  R-peak (T-wave).}

\begin{document}

\title{A room temperature 19-channel magnetic field mapping device for cardiac signals}

\def\JENA{Department of Neurology, Friedrich--Schiller--University, D--07740 Jena, Germany}
\def\UNIFR{Department of Physics, University of Fribourg, CH--1700, Fribourg, Switzerland}

\author{G. Bison}      \affiliation{\JENA}
\author{N. Castagna}   \affiliation{\UNIFR}
\author{A. Hofer}      \affiliation{\UNIFR}
\author{P. Knowles}    \affiliation{\UNIFR}
\author{J.-L. Schenker}\affiliation{\UNIFR}
\author{M. Kasprzak}   \affiliation{\UNIFR}
\author{H. Saudan}     \affiliation{\UNIFR}
\author{A. Weis}       \email[Corresponding author:~]{antoine.weis@unifr.ch} \affiliation{\UNIFR}

\date{\today}

\begin{abstract}
We present a multichannel cardiac magnetic field imaging system
built in Fribourg from optical double-resonance Cs vapor
magnetometers.
It consists of 25~individual sensors designed to record magnetic
field maps of the beating human heart by simultaneous measurements
on a grid of 19~points over the chest.
The system is operated as an array of second order
gradiometers using sophisticated digitally controlled feedback loops.
\end{abstract}

% no PACS for APL!
%\pacs{07.55.Ge, 32.30.Dx, 33.40.+f, 42.40.My, 42.62.-b, 87.57.-s}

\maketitle

%\section{Introduction}

Signals derived from the electrophysiological processes in the
human heart are important diagnostic tools for assessing cardiac
disorder.
For more than a century, measurements of body surface potentials
(electrocardiography, ECG) generated by electrical currents within
the myocardium have been a standard medical examination technique.
Electrophysiological currents also generate a magnetic field,
whose recording represents an alternative diagnostic tool \cite{Andra:1998:MM},
known as magnetocardiography (MCG).
Three decades after the first recording of an MCG signal \cite{Baule:1963:DMF}, Schneider \etal
\cite{Schneider:1990:MBS} in 1990 turned the MCG method into an imaging technique (cardiac magnetic field mapping,
cMFM) by using a sensor array covering the whole chest.
The magnetic signals (MCG) have a distinct advantage over the
electrical signals (ECG) when considering source localization: the
inverse problem of reconstructing the field sources from field maps is
easier to solve in the magnetic case since the electrical conductivity
of body tissue affects the transmission of magnetic signals to a much
lesser degree than their electric counterpart.
Modern cMFM systems \cite{BMDSys} use superconducting quantum interference devices (SQUIDs) for  recording the
very weak magnetic field (peak amplitude $\approx 100$~pT) generated by the heart.
In the past years we have developed laser optical pumping
magnetometers (LOPMs) \cite{Groeger:2006:HSL} for biomagnetic applications
\cite{Bison:2003:DMH}.
The room temperature operation of LOPMs makes such magnetometers
promising alternatives to SQUIDs by avoiding the expensive cryogenic
cooling and the associated complex logistics in view of clinical
applications.
Since each LOPM sensor requires only a few $\mu$W of light power, a
single laser can operate many dozens of individual sensor heads, a
feature which has until now been unexploited.
In this Letter we describe a cMFM system operating in Fribourg consisting of 25~individual LOPM sensors designed
to record magnetic field maps of the beating human heart by simultaneous measurements on a grid of 19~points over
the chest.

%\section{System design}

The LOPM sensors are based on optical-rf double resonance in the
so-called $M_x$ configuration \cite{Groeger:2006:HSL}.
The sensor medium is a room temperature Cs vapor contained in a
spherical paraffin-coated \cite{Castagna:2009:LSS} 30~mm diameter glass
bulb.
Circularly polarized light from a diode laser, frequency locked to the
$4{\rightarrow}3$ component of the cesium $D_1$ transition, creates a
macroscopic magnetization in the vapor by optical pumping.
The magnetization precesses around a static magnetic field $\vec{B}_{0} = \B \, \hat z $ --- nominally oriented at
$\pi/4$ with respect to the light propagation --- with the Larmor frequency
$\omega_L = \gamma \, | \B |$,
where $\gamma\approx 2 \pi \times 3.5~\mathrm{Hz}/\mathrm{nT}$.
The precession is coherently driven by a weak magnetic field
$\Brf\ll\B$ oscillating at the frequency \omegarf.
As a consequence, the light transmitted through the cell
acquires an amplitude modulation at the frequency
\omegarf.
The modulation amplitude is resonantly enhanced at $\omegarf =
\omega_L$ and the phase \Dfi{} between the drive and the response
obeys (up to a constant phase offset)
\begin{equation}
\Dfi=\arctan\left( \frac{\omega_L - \omegarf}{\Gamma} \right)\,,
\label{eqn.signal}
\end{equation}
where the (HWHM) linewidth $\Gamma$ is typically
$\approx 2 \pi \times 6~\mbox{Hz}$ \cite{Castagna:2009:LSS}
for our OPMs.
Near resonance, \Dfi{} depends linearly on \B\ and on \omegarf, so that feedback control of either \B{} or
\omegarf{} can be used to keep the system on resonance.

%\section{System Design}

The design of the 19-channel system was targeted at
demonstrating the simultaneous operation of many LOPM sensors in
close packing.
Each sensor consists of a compact self-contained module including
optical components to collimate and polarize the light, a Cs vapor
cell, and a photodiode to detect the light transmitted through the cell.
The components are mounted between two printed circuit boards
(PCB) which carry coils for generating \Brf.
Each sensor module is placed in an additional pair of coils
allowing the application of a small field, \Bfb, parallel to the
field \B.
These feedback coils are realized as printed circuits on
larger PCBs that serve as mechanical support structures
positioning the sensors in horizontal planes (Fig.~\ref{fig:setup}).
An optimal sensor packing is achieved by arranging the modules on
a hexagonal grid with a 50~mm spacing.
%
%Each module is connected to the control and acquisition
%electronics by miniature coaxial cables carrying the rf and
%photocurrents.
%
A custom-made hologram splits the beam from the
frequency-stabilized laser into 25 beams of equal intensity which
are carried by multimode optical fibers to the individual sensor
modules.
The complete sensor system consists of 25 modules distributed in
three planes that are vertically offset by 10~cm
(Fig.~\ref{fig:setup}).
The lower plane, closest to the subject's chest, is equipped with
19 primary mapping modules.
The middle and upper planes each carry three sensors providing
reference signals for stabilizing \B{} and its linear gradient,
respectively.
The sensor system is mounted inside of a two layer Al shielding room that reduces ambient magnetic field
fluctuation which are dominated by 50Hz (power line) and 16 2/3 Hz (railway) interference.
The person to be diagnosed is placed underneath the sensor system on a
nonmagnetic bed.
Three pairs of large coils surrounding the shielding room are used to
create a vertical field $\B\approx 9 \mu$T (corresponding to $\omega_L
= 2 \pi \times 31.5$~kHz) at the sensor location.
Transverse $d\B/dx$, $d\B/dy$ gradients are compensated using
additional large coils and constant current supplies.

\ifAPLsubmission
  % figures at end for submissions
\else
  \begin{figure}[t]
  \includegraphics*[width=\linewidth]{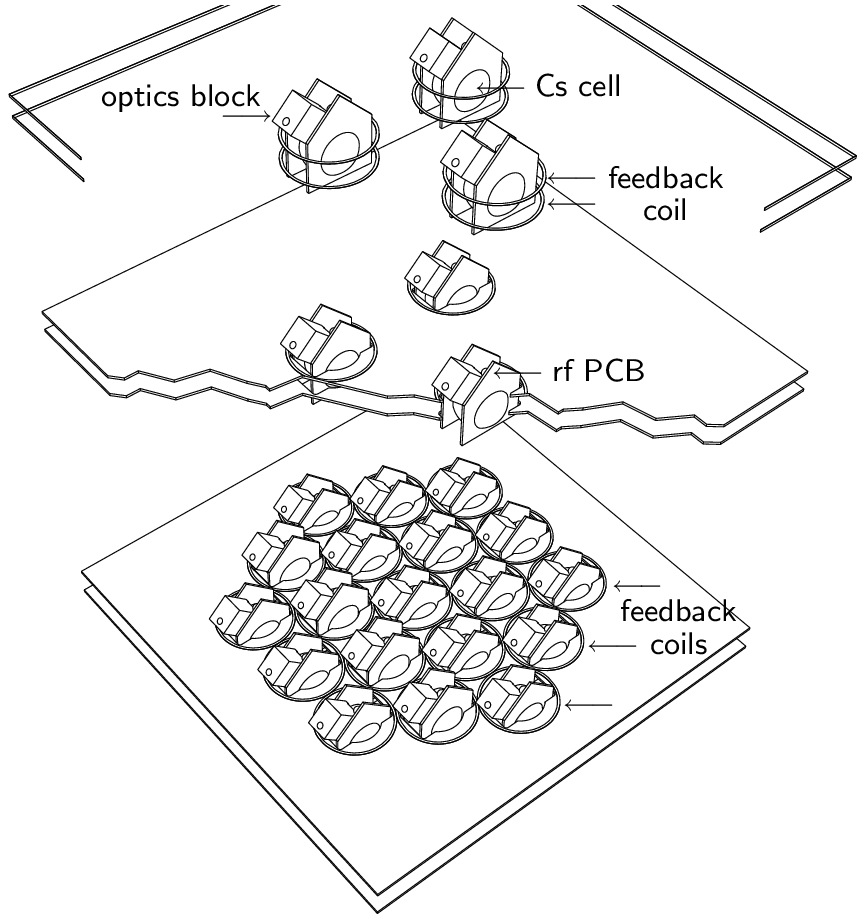}
  \caption{\Captionone}
  \label{fig:setup}
  \end{figure}
\fi

%\subsection{Electronics and offset field tracking}

The control  electronics is implemented in a digital FPGA board which digitizes, via ADCs, the photocurrent from
each module after transimpedance amplification, and which generates, via DACs, the currents driving the rf coils
and the feedback currents for the sensors in the lower and upper planes.
For each sensor the phase difference \Dfi{} is determined using a
numerical lock-in detection algorithm.
The near resonant linear dependence of the  phase signals form
suitable error signals that are used, after
proportional-integral-differential (PID) amplification, as
feedback signals in the various servo-loops discussed below.
The middle plane carries three sensors that serve as reference
sensors which track changes of the DC offset field \B{} (applied
field plus external field changes).
In each of the three middle plane sensors, \Dfi{} is actively
locked to zero by servo adjustments of the corresponding three drive
frequencies \omegarf{} generated by a numerically controlled
oscillator.
The computed average of those three individual frequencies or, alternatively, the frequency of a single (centered) sensor is
then used as common drive frequency \omegarfavg\ for the sensors in the top and bottom planes.
In both ways the drive frequency tracks fluctuations and drifts of \B.

%\subsection{First order gradient compensation}

When using a common drive frequency $\omegarfavg$, the signals \Dfi{} of the sensors in the top and bottom planes
are proportional to the difference between the field at their locations and the field in the middle plane.
These sensors thus form first-order hardware gradiometers.
In particular, the signal \Dfi{} of the top sensors measures the linear vertical magnetic field gradient, while
being almost insensitive to the cardiac field.
The phase differences \Dfi{} of the upper sensors are locked to zero
via servo loops injecting current into each of their respective
feedback coils.
In this way the local field at each upper sensor location follows the field in the middle plane.
When simultaneously injecting currents of the same magnitude, but of opposite polarity, into the bottom layer
feedback coils, the linear vertical field gradient along the vertical extension of the sensor system, and its
fluctuations are actively compensated.
Since the number of upper reference sensors is less than the number of
imaging sensors in the bottom plane, only the average of the upper
correction currents is used as feedback for the lower plane.
%
%As such, the odd-order gradients are cancelled to the level dictated
%by the finite geometry while the even-order gradient fluctuations
%remain, the summed effect of which are called here ``gradient effects''.

%\subsection{Mapping signals}

When the average field and the linear vertical gradient fluctuations are suppressed by the feedback loops
described above, the remaining phase differences \Dfi{} of the lower plane sensors are proportional to the second
order gradient of \B.
However, those phase differences are not directly measured: by using
yet another set of servo-loops driving the lower plane feedback coils,
the 19 phase differences \Dfi{} are removed, forcing all sensors to be
in resonance with \omegarfavg.
The 19 time-dependent feedback currents of the latter loop thus consist of the biomagnetic signals of interest and
fluctuations of \B's second order gradient.
These signals are recorded at a sampling rate of 5 kHz.
This magnetic feedback scheme relying on three distinct sets of
servo-loops is of general interest for any gradiometric arrangement of
multiple, closely packed magnetometers.
It may be particularly useful for coherent population trapping
magnetometers (CPTM), which have been used for biomagnetic field
detection~\cite{Belfi:2007:CCP} but have not yet been reported to be
scalable to many channels.
Close-packed CPTMs can be operated in an all-optical mode by using one
optical modulator per channel, however, magnetic feedback is probably
a more cost effective operation method for a multichannel CPTM
implementation.
The second order gradiometer scheme reduces field fluctuations by a factor of $>$1000 to a level of 900 \fTHz\ at 0.1
Hz.
The noise of the gradiometer signal drops towards larger frequencies and reaches 300 \fTHz\ at 100 Hz.

%\subsection{Sensor cross-talk}
Since the sensors are closely packed, the \Bfb{} field of a given
feedback coil induces crosstalk with its neighbours, although the
effect is small enough (${<}10\%$) that the system operates without
online corrections.
The measured crosstalk $dB_i/dI_j$ (field in sensor $i$ induced by
current in feedback coil $j$) can be used for offline data
correction\cite{Hofer:2009:HSO}.

\ifAPLsubmission
  % figures at end for submissions
\else
  \begin{figure}[t]
  \includegraphics*[width=\linewidth,angle=-90]{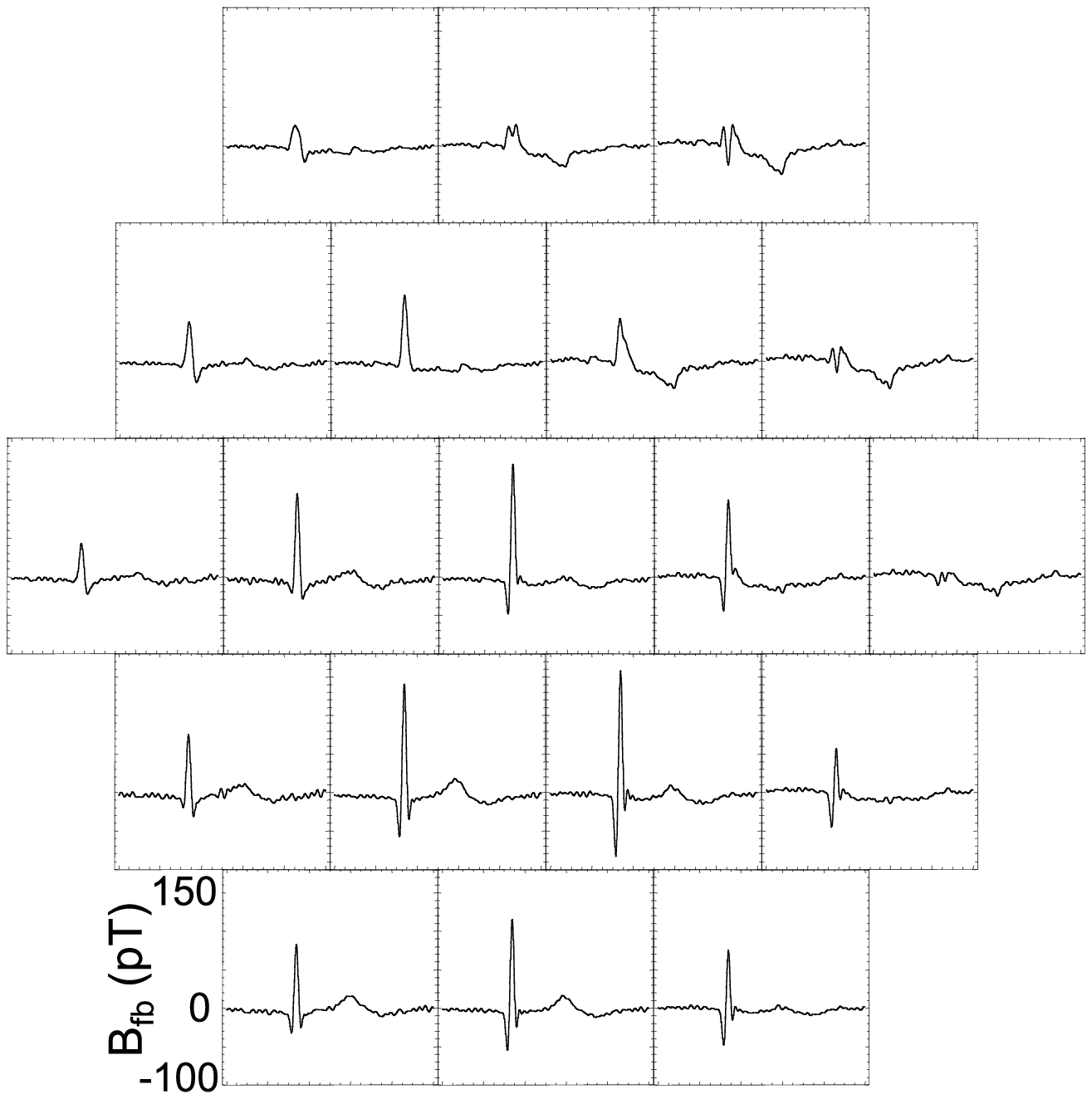}
  \caption{\Captiontwo}
  \label{fig:trace}
  \includegraphics*[width=\linewidth,angle=-0]{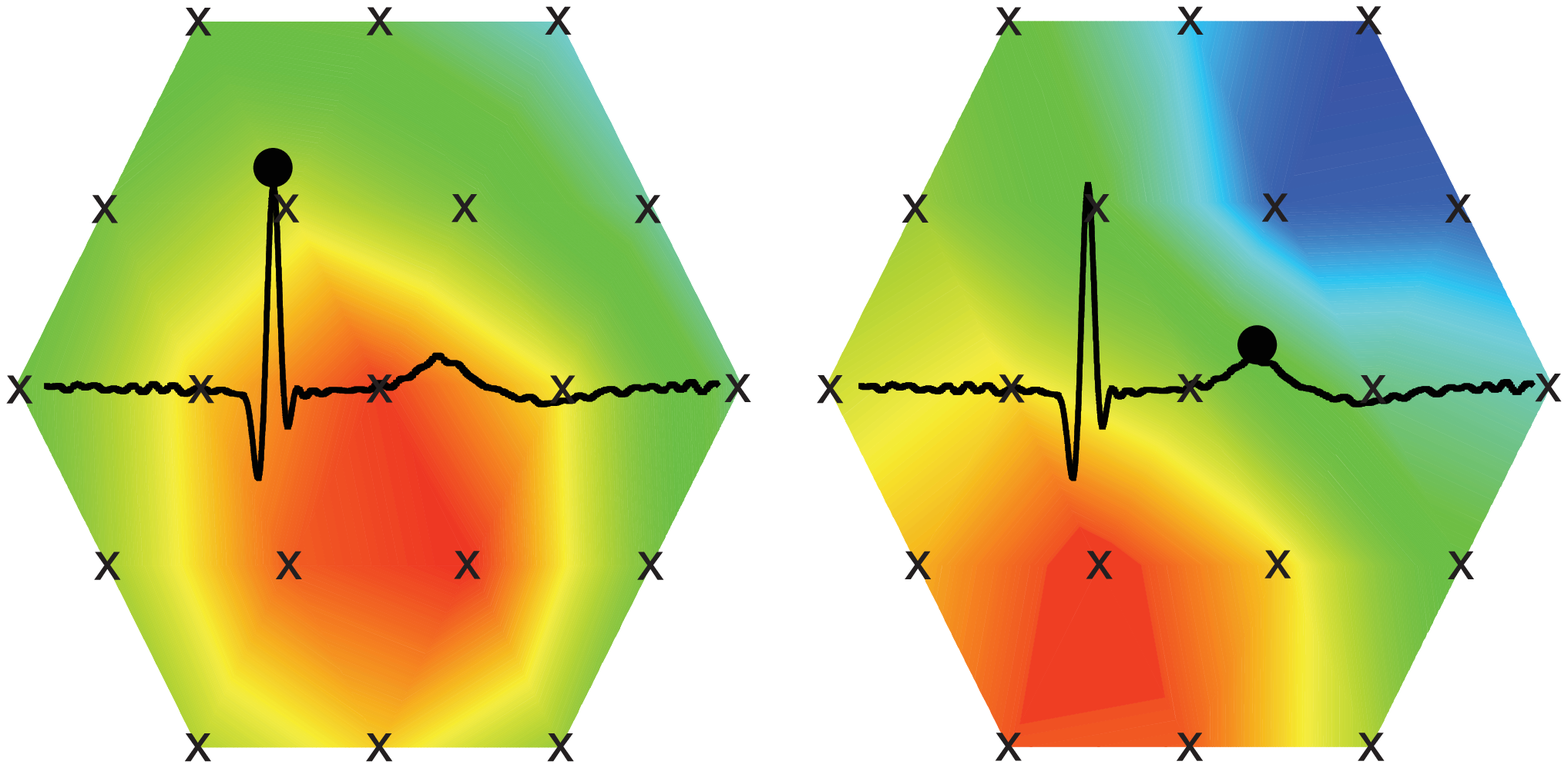}
  \caption{\Captionthree}
  \label{fig:map}
  \end{figure}
\fi

Figure~\ref{fig:trace} shows an example of 19 simultaneously recorded MCG traces obtained by averaging signals
from 66 heartbeats.
The data represents the second order gradient of the out-of-chest component of the cardiac field.
A 3-lead ECG signal served as timing reference for aligning the
individual heart beats when averaging.
The total recording time is less than two minutes, a huge improvement
compared to the two hours required when stepping a single sensor over
the chest \cite{Bison:2003:DMH}.
The signals are (low-pass) filtered offline by multiplying their Fourier spectra with a smooth function passing
from 1 to 0 in the band of 40 to 50 Hz.
Systematic implementation of crosstalk correction is a work in
progress and thus has not been applied to the shown data.
The residual oscillations are due to railway interference (16 2/3~Hz).
In Fig.~\ref{fig:map}, magnetic field intensity maps for two given
times (R-peak, T-wave) of the cardiac cycle are shown.
The smooth pattern is obtained by a numerical interpolation between the discrete measurement points
(shown as crosses).
%

%\section{Summary and Outlook}

We have reported the operation of a 19 channel second-order gradiometer system based on atomic magnetometers for
MCG measurements, capable of recording a magnetic field map in about two minutes in a weakly shielded environment.
The $19+3+3$ system is now being used at the University of Fribourg
for an initial series of measurements on both volunteers and cardiac
patients.
A $57+13+13$ channel system is currently being built at the University Hospital in Jena (Germany).
Tests of that system in a clinical environment should start early
in 2010.

\begin{acknowledgments}

This work was supported by the Velux Foundation,
% nEDM pays Catherine!
the Swiss National Science Foundation (\#200020--119820),
and the German Federal Ministry for Education and Research (\#0315024~B\@).
The authors thank Dr.~S.~N.~Ern{\'e} of BMDSys GmbH for help in the
sensor layout design.
\end{acknowledgments}

%\bibliographystyle{apsrev}
%\bibliography{FRAPref}

\begin{thebibliography}{9}
\expandafter\ifx\csname natexlab\endcsname\relax\def\natexlab#1{#1}\fi
\expandafter\ifx\csname bibnamefont\endcsname\relax
  \def\bibnamefont#1{#1}\fi
\expandafter\ifx\csname bibfnamefont\endcsname\relax
  \def\bibfnamefont#1{#1}\fi
\expandafter\ifx\csname citenamefont\endcsname\relax
  \def\citenamefont#1{#1}\fi
\expandafter\ifx\csname url\endcsname\relax
  \def\url#1{\texttt{#1}}\fi
\expandafter\ifx\csname urlprefix\endcsname\relax\def\urlprefix{URL }\fi
\providecommand{\bibinfo}[2]{#2}
\providecommand{\eprint}[2][]{\url{#2}}

\bibitem[{\citenamefont{Andr{\"{a}} and Nowak}(1998)}]{Andra:1998:MM}
\bibinfo{editor}{\bibfnamefont{W.}~\bibnamefont{Andr{\"{a}}}} \bibnamefont{and}
  \bibinfo{editor}{\bibfnamefont{H.}~\bibnamefont{Nowak}}, eds.,
  \emph{\bibinfo{title}{Magnetism in Medicine}} (\bibinfo{publisher}{Wiley--VCH
  Verlag GmbH}, \bibinfo{address}{Berlin}, \bibinfo{year}{1998}), ISBN
  \bibinfo{isbn}{3-527-40221-7}.

\bibitem[{\citenamefont{Baule and McFee}(1963)}]{Baule:1963:DMF}
\bibinfo{author}{\bibfnamefont{G.~M.} \bibnamefont{Baule}} \bibnamefont{and}
  \bibinfo{author}{\bibfnamefont{R.}~\bibnamefont{McFee}},
  \bibinfo{journal}{Am.\ Heart J.} \textbf{\bibinfo{volume}{66}},
  \bibinfo{pages}{95} (\bibinfo{year}{1963}).

\bibitem[{\citenamefont{Schneider et~al.}(1990)\citenamefont{Schneider, Hoenig,
  Reichenberger, Abraham-Fuchs, Moshage, Oppelt, Stefan, Weikl, and
  Wirth}}]{Schneider:1990:MBS}
\bibinfo{author}{\bibfnamefont{S.}~\bibnamefont{Schneider}},
  \bibinfo{author}{\bibfnamefont{E.}~\bibnamefont{Hoenig}},
  \bibinfo{author}{\bibfnamefont{H.}~\bibnamefont{Reichenberger}},
  \bibinfo{author}{\bibfnamefont{K.}~\bibnamefont{Abraham-Fuchs}},
  \bibinfo{author}{\bibfnamefont{W.}~\bibnamefont{Moshage}},
  \bibinfo{author}{\bibfnamefont{W.}~\bibnamefont{Oppelt}},
  \bibinfo{author}{\bibfnamefont{A.}~\bibnamefont{Stefan}},
  \bibinfo{author}{\bibfnamefont{A.}~\bibnamefont{Weikl}}, \bibnamefont{and}
  \bibinfo{author}{\bibfnamefont{A.}~\bibnamefont{Wirth}},
  \bibinfo{journal}{Radiology} \textbf{\bibinfo{volume}{176}},
  \bibinfo{pages}{825} (\bibinfo{year}{1990}).

\bibitem[{BMD()}]{BMDSys}
\bibinfo{note}{{BMDSys} {GmbH}, \url{http://www.bmdsys.com}}.

\bibitem[{\citenamefont{Groeger et~al.}(2006)\citenamefont{Groeger, Bison,
  Schenker, Wynands, and Weis}}]{Groeger:2006:HSL}
\bibinfo{author}{\bibfnamefont{S.}~\bibnamefont{Groeger}},
  \bibinfo{author}{\bibfnamefont{G.}~\bibnamefont{Bison}},
  \bibinfo{author}{\bibfnamefont{J.-L.} \bibnamefont{Schenker}},
  \bibinfo{author}{\bibfnamefont{R.}~\bibnamefont{Wynands}}, \bibnamefont{and}
  \bibinfo{author}{\bibfnamefont{A.}~\bibnamefont{Weis}},
  \bibinfo{journal}{Eur.~Phys.~J.~D} \textbf{\bibinfo{volume}{38}},
  \bibinfo{pages}{239} (\bibinfo{year}{2006}).

\bibitem[{\citenamefont{Bison et~al.}(2003)\citenamefont{Bison, Wynands, and
  Weis}}]{Bison:2003:DMH}
\bibinfo{author}{\bibfnamefont{G.}~\bibnamefont{Bison}},
  \bibinfo{author}{\bibfnamefont{R.}~\bibnamefont{Wynands}}, \bibnamefont{and}
  \bibinfo{author}{\bibfnamefont{A.}~\bibnamefont{Weis}},
  \bibinfo{journal}{Opt.~Expr.} \textbf{\bibinfo{volume}{11}},
  \bibinfo{pages}{904} (\bibinfo{year}{2003}).

\bibitem[{\citenamefont{Castagna et~al.}(2009)\citenamefont{Castagna, Bison,
  {Di~Domenico}, Hofer, Knowles, Macchione, Saudan, and
  Weis}}]{Castagna:2009:LSS}
\bibinfo{author}{\bibfnamefont{N.}~\bibnamefont{Castagna}},
  \bibinfo{author}{\bibfnamefont{G.}~\bibnamefont{Bison}},
  \bibinfo{author}{\bibfnamefont{G.}~\bibnamefont{{Di~Domenico}}},
  \bibinfo{author}{\bibfnamefont{A.}~\bibnamefont{Hofer}},
  \bibinfo{author}{\bibfnamefont{P.}~\bibnamefont{Knowles}},
  \bibinfo{author}{\bibfnamefont{C.}~\bibnamefont{Macchione}},
  \bibinfo{author}{\bibfnamefont{H.}~\bibnamefont{Saudan}}, \bibnamefont{and}
  \bibinfo{author}{\bibfnamefont{A.}~\bibnamefont{Weis}},
  \bibinfo{journal}{Appl.~Phys.~B}  \textbf{\bibinfo{volume}{96}},
  \bibinfo{pages}{763} (\bibinfo{year}{2009}).

\bibitem[{\citenamefont{Belfi et~al.}(2007)\citenamefont{Belfi, Bevilacqua,
  Biancalana, Cartaleva, Dancheva, and Moi}}]{Belfi:2007:CCP}
\bibinfo{author}{\bibfnamefont{J.}~\bibnamefont{Belfi}},
  \bibinfo{author}{\bibfnamefont{G.}~\bibnamefont{Bevilacqua}},
  \bibinfo{author}{\bibfnamefont{V.}~\bibnamefont{Biancalana}},
  \bibinfo{author}{\bibfnamefont{S.}~\bibnamefont{Cartaleva}},
  \bibinfo{author}{\bibfnamefont{Y.}~\bibnamefont{Dancheva}}, \bibnamefont{and}
  \bibinfo{author}{\bibfnamefont{L.}~\bibnamefont{Moi}}, \bibinfo{journal}{J.
  Opt. Soc. Am. B} \textbf{\bibinfo{volume}{24}}, \bibinfo{pages}{2357}
  (\bibinfo{year}{2007}).

\bibitem[{\citenamefont{Hofer et~al.}(2009)\citenamefont{Hofer, Bison,
  Castagna, Knowles, Schenker, and Weis}}]{Hofer:2009:HSO}
\bibinfo{author}{\bibfnamefont{A.}~\bibnamefont{Hofer}},
  \bibinfo{author}{\bibfnamefont{G.}~\bibnamefont{Bison}},
  \bibinfo{author}{\bibfnamefont{N.}~\bibnamefont{Castagna}},
  \bibinfo{author}{\bibfnamefont{P.}~\bibnamefont{Knowles}},
  \bibinfo{author}{\bibfnamefont{J.-L.} \bibnamefont{Schenker}},
  \bibnamefont{and} \bibinfo{author}{\bibfnamefont{A.}~\bibnamefont{Weis}},
  \bibinfo{journal}{unpublished}  (\bibinfo{year}{2009}).

\end{thebibliography}

\ifAPLsubmission
  %
  %List of figure captions is at end
  %
  \begin{figure}[b]
  \caption{\Captionone}
  \label{fig:setup}
  \caption{\Captiontwo}
  \label{fig:trace}
  \caption{\Captionthree}
  \label{fig:map}
  \end{figure}
\fi

\end{document}